\documentclass[showkeys,showpacs,amssymb,eqsecnum,superscriptaddress]{revtex4}

\renewcommand{\theequation}{\arabic{section}.\arabic{equation}}
\newcommand{\half}{\frac{1}{2}}
\newcommand{\bright}{\begin{flushright}}
\newcommand{\eright}{\end{flushright}}
\newcommand{\bminip}{\begin{minipage}}
\newcommand{\eminip}{\end{minipage}}
\newcommand{\bcent}{\begin{center}}
\newcommand{\ecent}{\end{center}}

\newcommand{\nnb}{\nonumber}
\newcommand{\reflef}{(\ref}
\newcommand{\MP}{M_{\rm P}}
\newcommand{\BBbox}{\Box}
\newcommand{\lmd}{\lambda}

\newcommand{\gsim}{\mbox{\raisebox{-.3em}{$\;\stackrel{>}{\sim}\;$}}}
\newcommand{\lsim}{\mbox{\raisebox{-.3em}{$\;\stackrel{<}{\sim}\;$}}}

\newcommand{\slashlmd}{$\lmd$\hspace{-.6em}\raisebox{.3em}{$-$}}
\newcommand{\beq}{\begin{equation}}
\newcommand{\eeq}{\end{equation}}
\newcommand{\beqa}{\begin{eqnarray}}
\newcommand{\eeqa}{\end{eqnarray}}
\newcommand{\bfig}{\begin{figure}}
\newcommand{\efig}{\end{figure}}

\begin{document}

\title{Gravitational scalar field coupled directly 
to the Maxwell field\\
 and its effect to solar-system experiments
} 
\author{Yasunori Fujii}
\affiliation{Advanced Research Institute for Science and Engineering,
Waseda University, Tokyo, 169-8555 Japan}
\author{Misao Sasaki}
\affiliation{Yukawa Institute for Theoretical Physics,
Kyoto University, Kyoto, 606-8502 Japan}

\begin{abstract}
The effect of the massless gravitational scalar field assumed to couple
 directly to the Maxwell field to the solar-system experiments is estimated. 
 We start with discussing the theoretical significances of this coupling.  
Rather disappointingly, however, we find that the scalar-field parameters never
 affect the observation in the limit of the geometric optics, indicating
 a marked difference from the well-known contribution through the
 spacetime metric.

\hfill \vbox{ \hbox{YITP-06-42}}
\end{abstract}


\maketitle
\begin{center}
\end{center}

\baselineskip=.52cm
\section{Introduction}

The light-ray passing near the Sun is subject to the time-delay of propagation proportional to $(1+\gamma)a$, where $\gamma$ is one of Eddington's parameters, while $a$ is Sun's Schwarzschild radius.  The most stringent test of General Relativity (GR) has been made by using the Cassini spacecraft with the measured result \cite{bert},
\beq
\gamma=1+ (2.1\pm 2.3)\times 10^{-5}.
\label{mn1-1}
\eeq
This is compared with the prediction from the Brans-Dicke model \cite{bd} of the scalar-tensor theory (STT) \cite{cup};
\beq
\gamma = 1-4\zeta^2,
\label{mn1-2}
\eeq
where
\beq
\zeta^{-2}=6+\epsilon \xi^{-1},\quad\mbox{with}\quad \epsilon =\pm 1,
\label{mn1-3}
\eeq
or also given by the Jordan-Brans-Dicke parameter 
$\omega_0=\epsilon (4\xi)^{-1}$ and $\epsilon = {\rm Sign} (\omega_0)$.
The result \reflef{mn1-1}) barely allows a negative region of
$\gamma -1$ extending to $-0.2\times 10^{-5}$.  This can be consistent
with \reflef{mn1-2}) if  $4\zeta^2 \lsim 0.2 \times 10^{-5}$,  translated into $\epsilon = +1$ and 
\beq
\xi \lsim 0.5\times 10^{-6},\quad\mbox{or}\quad \omega_0 \gsim 5\times 10^5.
\label{mn1-4}
\eeq
We might complain that the obtained size of $\xi$ is unnaturally small to the extent that it appears as if STT itself were nearly dying.

A possible way out is to assume the scalar field to mediate the force of a
 finite range,  refusing to reach the distance  relevant to the solar-system 
experiment in such a way that, as discussed in \cite{cup}, the experiment of 
the kind of \cite{bert} is no longer sensitive to the coupling strength between
the matter and the scalar field.  In the present study instead we continue to 
assume the massless field, but focus upon another type of the scalar-field 
interaction, the direct coupling to the electromagnetic field.  We emphasize 
that this coupling, if there is any, is outside the BD model mentioned above, 
hence not included in the estimate \reflef{mn1-2}).  Also the effect of this 
coupling to the solar-system experiment has never been explored seriously, 
as far as we know, yet appears tractable.  In spite of these alluring features,
 we eventually reach a negative conclusion of no chance to detect the effect
of the scalar field, as long as we cling to the geometric optics
approximation.  We still 
believe the analysis to contain interesting ingredients worth presenting in 
some detail.

In order to highlight the theoretical background of the direct coupling, 
we start with summarizing how the result \reflef{mn1-2}) was derived.  
Brans and Dicke assumed the validity of Weak Equivalence Principle (WEP) 
implemented by the decoupling of the scalar field from the matter Lagrangian 
$L_{\rm matt}$, resulting in  the geodesic equation  for any matter particle,
 including the photon \cite{bd}.  Stated in a more rigorous language, 
we assume the existence of a unique conformal frame (called BDCF) with 
$L_{\rm matt}$ endowed with the above decoupling property.  
In other conformal frames (CFs), the geodesic equation acquires a nonzero
right-hand side, which turns out to contain the scalar field in the way
common to all the particles independently of any specific properties of
individual particles.
For this reason the Universal Free-Fall, an expression of WEP, is maintained 
intact.  Their model thus provides with the ``metric theory" implying that the
effects of the scalar field occur only through the spacetime metric.  
The weak-field approximation in BDCF yields the result \reflef{mn1-2}).

On the other hand, there have been some arguments on the possible presence of
direct coupling of the scalar field to the Maxwell field, described by a
gauge-invariant Lagrangian given by
\beq
{\cal L}_{\rm smx} = -\sqrt{-g}\frac{1}{4}\Phi g^{\mu\rho}g^{\nu\sigma}F_{\mu\nu}F_{\rho\sigma},
\label{mn1-5}
\eeq
where $\Phi$ stands for some combination of  the scalar field, and 
$F_{\mu\nu}=\partial_\mu A_\nu -\partial_\nu A_\mu$ \cite{bek,barrow}.  
Since the electromagnetic field is part of the matter in the sense of GR, 
this term violates WEP through the terms involving the scalar field, against 
one of BD's requirements.   In fact, the matrix elements of \reflef{mn1-5}) 
estimated for the matter fields depend generally not only on the scalar field 
but also on the electric charge, obviously bringing about the contributions of 
the scalar field depending on whether a falling object contains electrically 
charged constituents or not, for example.  In this context the effect 
of \reflef{mn1-5}) is not represented through the spacetime metric, requiring 
separate analysis for the physical effects.

As another aspect the field $\Phi^{1/2}$ can be absorbed into the modified 
electromagnetic field $\tilde{A}_\mu$ defined by $\Phi^{1/2}A_\mu$ in such a 
way that the terms of the highest derivatives agree with the standard 
form \reflef{mn1-5}) but without $\Phi$-dependence.  This rescaling combined 
with the coupling with other more conventional current of charged fields, like
the electron, for example, will cause the rescaling of the electric charge 
$e$, thus the fine-structure constant depending on the spacetime coordinates 
through $\Phi$.  This motivated Bekenstein to propose his theoretical 
model \cite{bek}.  Many developments have followed in this direction also 
taking the cosmological evolution of the scalar field into 
account \cite{barrow}.   It does not appear, however, that serious efforts
have been made studying how the proposal is related to the more fundamental
STT.  The interaction \reflef{mn1-5}) might be only an effective coupling 
derived from a deeper origin.  We point out that the spacetime-dependent
fine-structure constant may follow also from STT, not necessarily due to
the coupling like \reflef{mn1-5}), as discussed in Chapter 6.1 of
\cite{cup} and \cite{yfplb}.  

Another important suggestion for the coupling as in \reflef{mn1-5}) comes 
from string theory.  The field equations of the bosonic closed string sector 
are derived from the Lagrangian in 26 dimensions, as shown
 in  \cite{callan}, particularly in Eq. (3.4.58) of \cite{GSW};
\beqa
{\cal L}_{\rm str}&=&\sqrt{-g}e^{-2\hat{\Phi} }
\left( \half R +2g^{\bar{\mu}\bar{\nu}} \partial_{\bar{\mu}}\hat{\Phi}
 \partial_{\bar{\nu}}\hat{\Phi} 
 -\frac{1}{12}H_{\bar{\mu}\bar{\nu}\bar{\lmd}}H^{\bar{\mu}\bar{\nu}\bar{\lmd}}
\right) 
\nnb\\
&=&\sqrt{-g}\left( \half\xi\phi^2 R
 -\half\epsilon g^{\bar{\mu}\bar{\nu}}\partial_{\bar{\mu}}\phi
 \partial_{\bar{\nu}}\phi  -\frac{1}{12}\xi\phi^2
 H_{\bar{\mu}\bar{\nu}\bar{\lmd}}H^{\bar{\mu}\bar{\nu}\bar{\lmd}}   \right),
\label{mn1-6}
\eeqa
where $\phi=2e^{-\hat{\Phi}}$ has been introduced in the second line 
with $\epsilon = -1$ and $\xi^{-1}=4$, or $\omega_0 =-1$.  Remarkably enough,
the first two terms in the second line look the same as in STT in 4 dimensions.
It even seems as if STT had been prepared for string theory invented decades
later.

This also suggests that string theory is formulated in what is called the
string CF,  which plays the same role as BDCF.  The last term in the 
parenthesis in the second line of \reflef{mn1-6}) then indicates that the 
gauge-field Lagrangian in 4 dimensions is likely to be multiplied by the 
scalar field, basically in the same way as in \reflef{mn1-5}).   
Although the conclusion depends on the yet-to-be-established details in the 
theoretical transition to the physical 4-dimensional spacetime, the coupling
 of this type and hence WEP violation appear rather generic \cite{dm}.  

As a concomitant aspect of the approach we notice the unmistakable 
sign $\epsilon =-1$ in \reflef{mn1-6}), implying a negative kinetic energy
 of $\phi$ according to our sign convention, though the mixing coupling 
between $\phi$ and the spinless portion of the metric tensor caused by the 
nonminimal coupling term recovers eventually the right sign corresponding to 
positive energies under the condition $\zeta^2 >0$, in agreement with the 
unitarity requirement imposed in deriving the field equations.  We add
that this positive-energy condition is met by the choice $\epsilon
\xi^{-1}=-4$ both in 4 and 26 dimensions.     
We point out that the same result $\epsilon =-1$ is shared by the dimensional 
compactification {\it a la} Kaluza-Klein (KK), as demonstrated in Eq. (1.23) 
of \cite{cup}, and is also favored by the cosmological equations in the
 presence of the cosmological constant, as elaborated in Chapter 4.4 
of \cite{cup}.  Further noteworthy is the likely occurrence of the direct
 interaction of the type \reflef{mn1-5}) in the KK compactification, as well.

{}From \reflef{mn1-3}) we further derive 
$0<\zeta^2 \hspace{.2em}\raisebox{.3em}{$<$}\hspace{-.7em}\raisebox{-.25em}{$>$}\hspace{.2em}1/6$ for $\epsilon =\pm 1$ \cite{bls}.  
In particular the scalar-field-matter decoupling assumed to play the central
 role in the Least Coupling Principle \cite{dp} is realized only for 
$\epsilon = +1$, contrary to the simple-minded interpretation of 
string theory, also to the KK approach, as well as to the cosmological constraint mentioned above.  We also note in this connection that no argument has been 
offered in  \cite{dp} for the suspected sign reversal in the string-theoretical process of descending to 4 dimensions.

Given the theoretical significances as we have learned above, observational
 consequences of \reflef{mn1-5}) appear to deserve further scrutiny whatever 
the origin.  We here focus upon the possible effect to the solar-system 
experiments.  The result may turn out to be as large as \reflef{mn1-2}).  
We may even hope that the two effects, from \reflef{mn1-2}) and
 \reflef{mn1-5}), conspire to nearly cancel each other leaving a small 
deviation of $\gamma$ from unity, as given by \reflef{mn1-1}), still allowing
 much larger and hence more natural value of $\xi$ than indicated 
by \reflef{mn1-4}), likely with $\epsilon =-1$.  With this wishful
 anticipation, we derive the field equations in Section 2, 
and develop the geometric optics approximation
in Section 3 to be applied to light-rays passing near the Sun.
By studying the solution in the limit of geometric optics, however,
 we find the result which is independent of the scalar-field parameters.  
Section 4 is devoted to concluding remarks.

Appendix A accommodates details of simple but lengthy calculations to 
derive \reflef{nxt-24}).

\setcounter{equation}{0}
\section{Field equations }

In order to have an idea on what the scalar field is expected to be like, 
we start with BD Lagrangian
\beq
{\cal L}_{\rm stt}=\sqrt{-g}\left( \half\xi \phi^2R
 -\half\epsilon g^{\mu\nu}\partial_\mu\phi\partial_\nu\phi +L_{\rm matter}
  \right),
\label{scfld1-1}
\eeq
where we use the reduced Planckian unit system with
 $c=\hbar=\MP \left( =\sqrt{c\hbar /8\pi G}\right) =1$.

We introduce the weak-field $\sigma$ for the scalar field by
\beq
\phi=\xi^{-1/2}\left(  1+\zeta\sigma\right),
\label{scfld1-4}
\eeq
where $\zeta$ is given by \reflef{mn1-3}).  We have chosen the parameters 
in such a way that the nonminimal coupling term, the first term in the 
parenthesis of \reflef{scfld1-1}), reduces to the standard Einstein-Hilbert 
term as $\sigma \rightarrow 0$.  After the process of diagonalization, 
we arrive at the field equation
\beq
\BBbox \sigma =\zeta T,
\label{scfld1-6}
\eeq
where $T$ is the trace of the matter energy-momentum tensor.

By adhering to the massless theory at this moment, the static field around 
the Sun is given by
\beq
\zeta\sigma \approx M_\odot \frac{\zeta^2}{4\pi r} = a\zeta^2 \frac{1}{r},
\label{scfld1-7}
\eeq
where the Schwarzschild radius of the Sun has been defined by
\beq
a=\frac{2M_\odot}{8\pi}=\frac{M_\odot}{4\pi}.
\label{scfld1-8}
\eeq

We note that \reflef{scfld1-6})  may not necessarily follow if $\sigma$ is 
interpreted  as the scalar field in \cite{bek}.  As a result the coefficients 
in \reflef{scfld1-7}) and \reflef{scfld1-8}) can be different.  
We nevertheless use the results here, the same in deriving \reflef{mn1-2}), 
because the final result does not depend on such details, as far as the scalar
force is assumed to be long-range, and the coupling strength is chosen to be 
of the gravitational size.

We also assume that \reflef{mn1-5}) reduces to the free Maxwell Lagrangian in
the limit $\sigma \rightarrow 0$.  In accordance with this we choose $\Phi$ to
be given by
\beq
\Phi =\left( \xi^{1/2}\phi \right)^\kappa
 =\left( 1+\zeta\sigma \right)^\kappa \approx 1+\zeta\kappa\sigma,
\label{scfld1-14}
\eeq
where $\kappa$ is a constant.  Accepting the scalar field as given 
by \reflef{scfld1-7})-\reflef{scfld1-14}) we now vary \reflef{mn1-5}) with
 respect to $A_\mu$ to derive
\beq
\nabla_\mu\left( \Phi F^{\mu\nu} \right)=0.
\label{scfld1-15_0}
\eeq
Ignoring contributions from other ordinary charged fields for the moment, 
it might be convenient to put \reflef{scfld1-15_0}) into the form
\beq
\nabla_\mu F^{\mu\nu}=-j^\nu = -\Psi_\mu F^{\mu\nu},
 \quad\mbox{with}\quad \Psi_\mu =\frac{\partial_\mu \Phi}{\Phi}.
\label{scfld1-15_5}
\eeq
By the repeated use of this equation, we readily derive the conservation law
\beq
\nabla_\nu j^\nu =0.
\label{scfld1-15_6}
\eeq

Basically in the same way as in the conventional electrodynamics, 
this conservation law allows us to impose the gauge condition
\beq
\chi =\nabla_\mu \left( \Phi^n A^\mu \right) =0,
\label{scfld1-15_7}
\eeq
where $n$ is an arbitrary real number.  Consistency of this condition with the
field equation is assured by the linear differential equation for $\chi$,
\beq
\BBbox \chi + n\Psi_\mu \left( \nabla^\mu \chi \right)=0,
\label{scfld1-15_7_8}
\eeq
derived also from the field equation, where $\BBbox =\nabla_\mu \nabla^\mu$.

By imposing \reflef{scfld1-15_7}), the field equation \reflef{scfld1-15_0}) is 
put into 
\beq
\BBbox A^\mu +\Psi_\nu \left( \raisebox{-.1em}{\rule{0em}{1.1em}}
 \nabla^\mu A^\nu -(1-n) \nabla^\nu A^\mu \right)
 +n\left( \nabla^\mu \Psi^\nu \right) A_\nu + R^{\mu\nu}A_\nu=0,
\label{scfld1-15_8}
\eeq
where the last term comes from the failure of commutativity between 
$\nabla_\mu$ and $\nabla_\nu$ which we encountered in re-arranging terms in 
the far-left-hand side of \reflef{scfld1-15_5}).

\setcounter{equation}{0}
\section{Geometric optics}

Let us first consider the geometric optics for the
conventional free Maxwell field with $\Phi =1$ \cite{Misner:1974qy};
\beq
\nabla_\mu F^{\mu\nu}=0.
\label{nxt-1}
\eeq
We put
\beq
A^\mu = a^\mu e^{iS},
\label{nxt-2}
\eeq
in which the phase $S$ is assumed to vary much faster than the amplitude
 $a^\mu$ does.   We also introduce the scalar amplitude ${\cal A}$ and the 
normalized polarization vector $\epsilon^\mu$ in such a way
\beq
a^\mu ={\cal A}\epsilon^\mu, \quad\mbox{and}\quad \epsilon_\mu \epsilon^\mu =1.
\label{nxt-3}
\eeq

We find
\beqa
F^{\mu\nu}
&=&\nabla^\mu \left( a^\nu  e^{iS}\right)
 - \nabla^\nu \left( a^\mu  e^{iS}\right)
\nnb\\
&=&\left[  i\left( k^\mu a^\nu - k^\nu a^\mu\right)
  +\left( \nabla^\mu a^\nu - \nabla^\nu a^\mu  \right) \right]e^{iS},
\label{nxt-4}
\eeqa
where
\beq
k^\mu =\nabla^\mu S,
\label{nxt-5}
\eeq
representing the normal to the surface of a constant phase, $S$, hence to be
called a wave-vector.  The first set of the two terms in the last line
 of \reflef{nxt-4}) comes from differentiating the phase $S$, hence to be 
called the terms of rank 1.  The two terms in the second set, on the other
 hand, are called the terms of rank 0 because no differentiation of $S$ is
 involved.  The rank $r$ expresses how many times $e^{iS}$
 is differentiated,
 corresponding to expanding $e^{iS/\epsilon}$ into the series of
 $\epsilon^{-r}$ in many of conventional calculations.
 A decrease of $r$ by every unit implies a factor as small 
as $\sim$ \hspace{.0em}\slashlmd$/\ell \sim 10^{-12}$,
 where \slashlmd\hspace{.1em} and $\ell$ are the wavelength of radiation and
 the  typical size of the solar system, respectively.

With the help of $k^\mu$, we impose a gauge condition
\beq
k_\mu \epsilon^\mu =0,
\label{nxt-6}
\eeq
corresponding to the choice $n=0$ in \reflef{scfld1-15_7}). 
Other choices of $n$ will be discussed later. 

Now we substitute \reflef{nxt-4}) into \reflef{nxt-1}), obtaining
\beqa
\hspace*{-.5em}\nabla_\mu F^{\mu\nu}&\hspace{-.9em}=&\hspace{-.6em}
 \left(
\raisebox{-.1em}{\rule{0em}{1.2em}}\hspace{-.2em}
ik_\mu \left[ i \left(k^\mu a^\nu - k^\nu a^\mu\right)
 +\left( \nabla^\mu a^\nu -\nabla^\nu a^\mu \right) \right] 
+ \nabla_\mu \left[  i \left(k^\mu a^\nu - k^\nu a^\mu\right)
 +\left( \nabla^\mu a^\nu -\nabla^\nu a^\mu \right) \right] \right) e^{iS}
\nnb\\
&\hspace{-.9em}=&\hspace{-.6em}
\left( \raisebox{-.1em}{\rule{0em}{1.2em}}\hspace{-.2em}
-k_\mu k^\mu a^\nu +k^\nu k_\mu a^\mu +i{\cal B}^\nu
  + {\cal C}^\nu\right) e^{iS},
\label{nxt-7}
\eeqa
where \reflef{nxt-6}) applies to the second term in the last line, and 
\beqa
{\cal B}^\nu &=&a^\nu\nabla_\mu k^\mu -a^\mu\nabla_\mu k^\nu
 +k^\mu \nabla_\mu a^\nu -k^\nu \nabla_\mu a^\mu
 +k_\mu \nabla^\mu a^\nu -k_\mu \nabla^\nu a^\mu,
\label{nxt-8}
\\
{\cal C}^\nu &=&\nabla_\mu \left( \nabla^\mu a^\nu
 - \nabla^\nu a^\mu \right)+R^\nu_{\hspace{.3em}\mu}a^\mu.
\label{nxt-8_1}
\eeqa
Note that the first and the second terms in the last line of \reflef{nxt-7})
 are of rank 2 (bilinear in $k$),  while ${\cal B}^\nu$ and ${\cal C}^\nu$
 have rank 1 and 0 (linear in $k$ and constant), respectively. 
 We have used \reflef{nxt-5}), also understanding that $\nabla_\mu$,
 for example, no  longer operates to $e^{iS}$, though
$\nabla_\mu (k^\mu a^\nu)=(\nabla_\mu k^\mu)a^\nu+k^\mu (\nabla_\mu a^\nu)$,
for example.  

We multiply \reflef{nxt-8}) by $\epsilon_\nu$.
Analyzing each term separately, as shown explicitly in Appendix A, we obtain
\beq
\epsilon_\nu {\cal B}^\nu ={\cal A}\left( \nabla_\mu k^\mu \right)
 +2k^\mu \left( \nabla_\mu {\cal A} \right).
\label{nxt-24}
\eeq

In order  to include the effect of the scalar field, we now go back
to \reflef{scfld1-15_0}) in its complete expression.  
By multiplying by $\epsilon_\nu$ we find
\beq
\Phi \epsilon_\nu\nabla_\mu  F^{\mu\nu}
 + \left( \nabla_\mu \Phi \right) \epsilon_\nu F^{\mu\nu}=0.
\label{nxt-32}
\eeq
For the first term we use \reflef{nxt-7}) and \reflef{nxt-24}) obtaining
\beq
\Phi \epsilon_\nu \left( -k_\mu k^\mu a^\nu +i{\cal B}^\nu
 + {\cal C}^\nu \right)e^{iS}
 =\Phi \left( \raisebox{-.1em}{\rule{0em}{1.2em}} -k_\mu k^\mu {\cal A}
  +i\left[ {\cal A}\left( \nabla_\mu k^\mu \right)
 + 2k^\mu  \nabla_\mu {\cal A}  \right] 
+ \epsilon_\nu{\cal C}^\nu \right)e^{iS}.
\label{nxt-33}
\eeq
For the second term in \reflef{nxt-32}) we use \reflef{nxt-4}) finding
\beq
\left( \nabla_\mu \Phi \right)\epsilon_\nu
 \left( \raisebox{-.1em}{\rule{0em}{1.2em}}
i\left( k^\mu a^\nu - k^\nu a^\mu\right) 
 +\left( \nabla^\mu a^\nu - \nabla^\nu a^\mu\right)   \right) e^{iS}
=  \left( \raisebox{-.1em}{\rule{0em}{1.2em}}
 i\left( \nabla_\mu \Phi \right) k^\mu {\cal A}
 - \epsilon_\nu
  \left( \nabla_\mu \Phi \right) \left( \nabla^\nu a^\mu \right)  \right)e^{iS},\label{nxt-34}
\eeq
where we have used 
\beq
\epsilon_\nu \nabla^\mu a^\nu
 =\nabla ^\mu\left( \epsilon_\nu a^\nu\right)
 -\left( \nabla^\nu \epsilon_\nu \right) a^\nu =0,
\label{nxt-34_1}
\eeq
due to \reflef{nxt-6}) and \reflef{s1-16}).  

Summing \reflef{nxt-33}) and \reflef{nxt-34}), we finally obtain
\beq
-\left( \Phi{\cal A} k_\mu k^\mu -2i {\cal F} -2{\cal G}  \right) e^{iS}=0,
\label{nxt-35}
\eeq
where
\beqa
{\cal F}&=&\Phi k^\mu \nabla_\mu {\cal A}
+\half {\cal A}\nabla_\mu \left( \Phi k^\mu \right),
\label{nxt-35_1} \\
{\cal G}&=& \Phi \epsilon_\nu {\cal C}^\nu
 -\left( \nabla_\mu \Phi \right) \epsilon_\nu \nabla^\nu a^\mu,
\label{nxt-35_1_1} 
\eeqa
which are real-valued, carrying the rank 1 and rank 0, respectively.

The complex-valued equation \reflef{nxt-35}) represents
 two real-valued equations
\beqa
k_\mu k^\mu &=&0,
\label{mn1_101}\\
{\cal F}&=&0,
\label{mn1_102}
\eeqa
where we have dropped $2{\cal G}/(\Phi{\cal A})$
  on the right-hand side of \reflef{mn1_101}) because it has rank $r=0$.

The first one \reflef{mn1_101}) implies a geodesic equation,
 based on the standard relation
\beq
\nabla_\nu\left( k_\mu k^\mu \right)=2k^\mu \left( \nabla_\nu k_\mu \right)
 =2k^\mu \nabla_\nu\left(\frac{\partial S}{\partial x^\mu}\right)
= 2k^\mu \left(\nabla_\mu k_\nu \right)
 = 2\frac{d x^\mu}{d \lmd}\left( \nabla_\mu k_\nu \right)
=2\frac{Dk_\nu}{D\lmd},
\label{nxt-36}
\eeq
where $\lmd$ is the distance measured along a ray defined by
\beq
k^\mu=\frac{d x^\mu}{d \lmd}.
\label{nxt-36-1}
\eeq
By accepting \reflef{mn1_101}), the far-left-hand side of \reflef{nxt-36})
 vanishes, and so does the far-right-hand side to result in
\beq
\frac{Dk_\nu}{D\lmd}=0.
\label{mn1_103}
\eeq
This tells us simply that the light-ray propagates exactly along the same 
geodesic as the one without the scalar field included.  In other words,
 the solar-system experiments using the light-rays fail to constrain the
 scalar-field parameters which describe how it couples to the Maxwell field
 through the direct interaction as in \reflef{mn1-5}).  
This might sound rather disappointing because the 
interaction \reflef{mn1-5}) breaks WEP, indicating the occurrence of the 
inhomogeneous term on the right-hand side of the geodesic equation. 
 We might be content with an interpretation that the effect fails to 
show up in the limit of the geometric optics.  

In fact on the right-hand side of \reflef{mn1_101}) we could have retained
the term of ${\cal G}/(\Phi{\cal A})$, which depends on the derivative of
$\Phi$ according to \reflef{nxt-35_1_1}), thus providing the inhomogeneous 
term we had expected.  Including such terms of the lower rank is hardly 
promising, however, in the realistic situation where the phenomena are 
described successfully by the geometric optics.

On the other hand, \reflef{mn1_102}) tells us how the amplitude of the ray 
is affected by the presence of the scalar field.  In this connection we first
notice that this equation in the absence of the scalar field, $\Phi =1$, 
reduces to
\beq
{\cal F}_0({\cal A}) =k^\mu \nabla_\mu {\cal A}
+\half {\cal A}\nabla_\mu \left(  k^\mu \right) =0,
\label{mn1_105}
\eeq
which is known to entail the conservation of photon flux \cite{Misner:1974qy}, ${\cal A}$ falling off
 like $({\rm distance})^{-1}$ in its propagation in a spherically symmetric
 flat spacetime, as shown toward the end of Appendix A.  In the presence of 
the scalar field, we find that \reflef{mn1_102}) given by \reflef{nxt-35_1})
 is re-expressed as
\beq
{\cal F}=\Phi^{1/2}{\cal F}_0({\cal A}_*) =0,
\label{mn1_107}
\eeq
in terms of the modified amplitude ${\cal A}_*$ defined by
\beq
{\cal A}=\Phi^{-1/2}{\cal A}_*.
\label{mn1_106}
\eeq

We could first calculate ${\cal A}_*$ as the photon-flux-conserving amplitude in 
Schwarzschild spacetime, and then use \reflef{mn1_106}) to obtain ${\cal A}$, 
representing how $\Phi$ affects the amplitudes, though no observational
 result is available at present.

\setcounter{equation}{0}
\section{Concluding remarks}

According to the result in the preceding section, the scalar-field parameters,
 typically $\kappa$ in \reflef{scfld1-14}), remain largely unconstrained.  
Probably we may look for other phenomena in which the effects will show up, 
like the one suggested in 6.5 of \cite{cup}.  At this moment, in particular, 
we admit disappointingly that we no longer have a reasonable candidate
 for producing $\gamma >1$ as indicated by the measurement \cite{bert}.
 The strong argument for this result \cite{bert2} may call for something entirely new.   Even in more general terms we failed to offer a successful scenario of conspiracy to save the simple-minded STT, by which $\gamma -1=-4\zeta^2$ in \reflef{mn1-2}) is nearly canceled by the 
contribution from \reflef{mn1-5}), thus allowing much larger and more natural 
value  of $\zeta^2>1/6$ hence of $\xi>1/6$ with $\epsilon =-1$ \cite{bls}.  
We might be inclined to support the idea of a massive scalar field \cite{cup}. 
See \cite{yfptp} for the detailed argument on the consistency between the 
massive and the massless behaviors of the scalar field in the local and the 
cosmological environments, respectively. 

As already mentioned in Section 1, there is another way to describe the 
Maxwell field by absorbing $\Phi$ into the electromagnetic field, which should
be important to discuss possible spacetime-dependent fine-structure constant. 
In fact \reflef{mn1-5}) is re-expressed as
\beq
{\cal L}_{\rm smx}
 = -\sqrt{-g}\frac{1}{4}
 g^{\mu\rho}g^{\nu\sigma}\tilde{F}_{\mu\nu}\tilde{F}_{\rho\sigma},
\label{mn1_151}
\eeq
where
\beq
\tilde{F}_{\mu\nu}
=\hat{F}_{\mu\nu} -\half \left( \Psi_\mu \tilde{A}_\nu
 - \Psi_\nu \tilde{A}_\mu \right),
\label{mn1_152}
\eeq
with
\beqa
A_\mu = \Phi^{-1/2}\tilde{A}_\mu,\quad\mbox{and}\quad 
\hat{F}_{\mu\nu} = \partial_\mu \tilde{A}_\nu - \partial_\nu \tilde{A}_\mu.
\label{mn1_153a}
\eeqa

The field equation with respect to $\tilde{A}_\mu$ as an independent
variable will be somewhat complicated because $\tilde{F}_{\rho\sigma}$ now 
depends on the un-differentiated $\tilde{A}_\rho$ even in flat spacetime as 
illustrated by
\beq
\frac{\partial \tilde{F}_{\rho\sigma}}{\partial \tilde{A}_\mu}
=-\half \left( \raisebox{-.1em}{\rule{0em}{1.2em}}\delta^\mu _{\sigma}
 \Psi_\rho - (\rho \leftrightarrow \sigma)\right).
\label{chsm_2} 
\eeq
Let us present the following explicit calculation only in flat spacetime,
for the moment, with understanding the ``comma-goes-to-semicolon'' rule \cite{Misner:1974qy}, except for 
the additional term of $R^\nu_{\hspace{.3em}\mu}A^\mu$, as we encountered 
in \reflef{scfld1-15_8}) and \reflef{nxt-8_1}).  From \reflef{mn1_151})
 we then derive
\beq
\left( \partial_\nu +\half \Psi_\nu \right)\tilde{F}^{\nu\mu} =0.
\label{renmx_12a}
\eeq

We impose the gauge condition
\beq
\tilde{\chi}=\partial_\mu \left( \Phi^{\tilde{n}} \tilde{A}^\mu \right) =0,
\label{chsm_1}
\eeq
which corresponds to choosing $n=\tilde{n}+1/2$ in \reflef{scfld1-15_7}).
 Due to this condition we have
\beq
\partial_\mu \tilde{A}^\mu =-\tilde{n} \Psi_\mu \tilde{A}^\mu.
\label{chsm_10_1}
\eeq
Substituting this into \reflef{renmx_12a}) together with \reflef{mn1_152})
 yields
\beq
\BBbox \tilde{A}^\mu +\left(\tilde{n}- \half \right)
 \Psi_\nu \left( \partial^\mu \tilde{A}^\nu \right)
+ \left( \raisebox{-.1em}{\rule{0em}{1.1em}}\mbox{terms of } r=0 \right)=0.
\label{chsm_9}
\eeq
Remarkably, the explicit scalar-field dependence appears
only in the terms of rank $r=0$ if $\tilde{n}=1/2$. 

We also note that the corresponding field equation for $A_\mu$ given
by \reflef{scfld1-15_8}) reduces to
\beq
\BBbox A^\mu
 + \left( \raisebox{-.1em}{\rule{0em}{1.1em}}\mbox{terms of } r=0 \right)=0,
\label{chsm_9a}
\eeq
if no scalar field is present.  We thus find that the field equation with 
respect to $\tilde{A}_\mu$ with the scalar field included but with the special 
choice $\tilde{n}=1/2$ is equivalent to the field equation with respect 
to $A_\mu$ without the scalar field, as far as we confine ourselves to the 
terms of $r=2$ and $r=1$.  In view of the gauge invariance of any of the
 physical observables, we come to conclude that the description in terms of 
$\tilde{A}_\mu$ is essentially the scalar-field-free description in
terms of $A_\mu$ for any 
sensible physical situation in which geometric optics applies including the 
amplitude variation corresponding to $r=1$.  

We close the paper by adding another comment: We carried out our
analysis exclusively in the CF identified with BDCF, 
though we are outside the pure 
BD model by introducing the WEP violating interaction.  As emphasized 
in 4.4.3 of \cite{cup}, however, this CF is most likely different from the 
physical CF in which we should expect time-independent masses of particles,
also with acceptable cosmological evolution in the presence of a 
cosmological constant.  In moving to the physical CF, we apply a conformal 
transformation with the function basically behaving like $\Omega \sim t^\eta$,
with $\eta$ a constant of the order one.  

We find, on the other hand, that the physically observed time-delay is 
obtained by spatially integrating the ratio
\beq
\frac{dt}{dr}= \frac{dt/d\lmd}{dr/d\lmd} =\frac{k^t}{k^r}.
\label{man1-92}
\eeq
It also follows that this ratio will be transformed roughly in the same 
manner as for $k^\nu$, resulting in a multiplicative factor $f=1+ \eta (\Delta t/t_0)$, 
where $t_0 \sim 10^{10} {\rm y} \sim 10^{17}{\rm sec}$ while 
$\Delta t \sim 10^3{\rm sec}$ for the approximate travel time of radiation in 
the solar system, yielding $f\sim 1$ to the accuracy of $\sim 10^{-14}$.  

The same argument applies also to the deflection of light, for which the 
ratio \reflef{man1-92}) is replaced by $d\phi/dr=k^\phi/k^r$. 
 
\acknowledgements
YF is grateful to Bruno Bertotti for his detailed account on his own
efforts and his strong interest in the direct coupling, which motivated
the present work.  He also thanks Minoru Omote for his many useful
discussions.  The work of MS is supported in part by JSPS Grant-in-Aid for Scientific
 Research (S) No.~14102004, (B) No.~17340075, and  (A) No.~18204024.

\appendix
\renewcommand{\thesection}{\Alph{section}}
\renewcommand{\theequation}{\Alph{section}.\arabic{equation}}
\setcounter{equation}{0}
\section{Deriving \reflef{nxt-24})}

Consider \reflef{nxt-8}),
\beq
{\cal B}^\nu= a^\nu \nabla_\mu k^\mu -a^\mu\nabla_\mu k^\nu 
+ k^\mu\nabla_\mu a^\nu -k^\nu \nabla_\mu a^\mu 
+ k_\mu \left( \nabla^\mu a^\nu - \nabla^\nu a^\mu\right).
\label{s1-1}
\eeq
Multiply with $\epsilon_\nu$ obtaining
\beqa
\epsilon_\nu {\cal B}^\nu
&=&(\epsilon \cdot a)\nabla_\mu k^\mu   \label{s1-2} \\
&-& a^\mu \epsilon_\nu \nabla_\mu k^\nu \label{s1-3} \\
&+& k^ \mu\epsilon_\nu\nabla_\mu a^\nu \label{s1-4} \\
&-& (\epsilon\cdot k)\nabla_\mu a^\mu \label{s1-5} \\
&+& k_\mu \epsilon_\nu \nabla^\mu a^\nu \label{s1-6} \\
&-& k_\mu \epsilon_\nu \nabla^\nu a^\mu. \label{s1-7} 
\eeqa
We find immediately
\beqa
{\rm \reflef{s1-2})}&=&{\cal A}\nabla_\mu k^\mu, \label{s1-8} \\
{\rm \reflef{s1-5})}&=& 0,\label{s1-9}
\eeqa
also with
\beqa
{\rm \reflef{s1-3})}
&=&-a^\mu\left[ \nabla_\mu \left( \epsilon\cdot k \right)
 -k^\nu \left( \nabla_\mu \epsilon_\nu \right)  \right]
 = a^\mu k^\nu \left( \nabla_\mu \epsilon_\nu \right),
  \label{s1-10} \\
{\rm \reflef{s1-4})}
&=& k^\mu \left[ \nabla_\mu (\epsilon\cdot a)
-a^\nu \left( \nabla_\mu\epsilon_\nu \right) \right]
 =k^\mu \nabla_\mu {\cal A} -k^\mu a^\nu \left( \nabla_\mu\epsilon_\nu \right),
 \label{s1-11} \\
{\rm \reflef{s1-6})}
&=&k_\mu \left[ \nabla^\mu (\epsilon\cdot a)
-a^\nu \nabla^\mu \left( \epsilon_\nu \right)\right]
 =k^\mu \nabla_\mu {\cal A}
 -k^\mu a^\nu \left( \nabla^\mu  \epsilon_\nu \right),
 \label{s1-12} \\
{\rm \reflef{s1-7})}
&=&-k_\nu \epsilon_\mu \left( \nabla^\mu a^\nu\right)
 =-k_\nu \epsilon_\mu \left[ \left( \nabla^\mu \epsilon^\nu \right){\cal A}
+\epsilon^\nu \left( \nabla^\mu {\cal A} \right)  \right],
 \nnb\\
&=&-{\cal A}k^\nu \epsilon^\mu \left( \nabla_\mu \epsilon_\nu \right)
 -(k\cdot \epsilon) \epsilon_\mu \nabla^\mu {\cal A}
 =-a^\mu k^\nu\left( \nabla_\mu \epsilon_\nu \right).
 \label{s1-13}
\eeqa
Collecting them we find
\beq
\epsilon_\nu {\cal B}^\nu ={\cal A}\left( \nabla_\mu k^\mu \right)
 +2k^\mu \left( \nabla_\mu {\cal A} \right) +{\cal R},
\label{s1-14}
\eeq
where the remainder is given by
\beq
{\cal R}= {\rm \reflef{s1-10})}+{\rm \reflef{s1-11})}
+{\rm \reflef{s1-12})}+{\rm \reflef{s1-13})}
=-2k^\mu a^\nu \left( \nabla_\mu \epsilon_\nu \right),
\label{s1-15}
\eeq
which vanishes  because
\beq
a^\nu \left( \nabla_\mu \epsilon_\nu \right)
 ={\cal A}\epsilon^\nu\left( \nabla_\mu \epsilon_\nu \right)
 =\frac{{\cal A}}{2}\left( \nabla_\mu \left(\epsilon\cdot \epsilon \right)
 \right) =\frac{{\cal A}}{2}\partial_\mu (1)=0.
\label{s1-16}
\eeq
We thus obtain  \reflef{nxt-24});
\beq
\epsilon_\nu {\cal B}^\nu = {\cal A}\left( \nabla_\mu k^\mu \right)
 +2k^\mu \left( \nabla_\mu {\cal A} \right).
\label{s1-16_1}
\eeq

We add that \reflef{s1-16_1}) allows a natural solution of
 ${\cal A}\propto r^{-1}$.  Consider spherically symmetric 3-space,
 for which we have
\beq
\left( \nabla_\mu k^\mu \right)
 =\left( \partial_\theta + \cot\theta \right)k^\theta
 +\partial_\phi k^\theta + \left( \partial_r+\frac{2}{r} \right)k^r.
\label{s1-22}
\eeq
Make simplifying assumptions
\beq
k^\theta = k^\phi =0,\quad\mbox{and}\quad \partial_r k^r\approx 0,
\label{s1-23}
\eeq
to put \reflef{s1-16_1}) into
\beq
{\cal A}\frac{2}{r}k^r +2k^r \frac{d{\cal A}}{dr}\approx 0,
\label{s1-24}
\eeq
finding
\beq
{\cal A}(r) \propto r^{-1}.
\label{s1-25}
\eeq

\end{document}